\documentclass{elsart}
\usepackage{epsfig}

\def\be{\begin{eqnarray}}
\def\ee{\end{eqnarray}}
\def\lsim{\mathrel{\rlap{\lower3pt\hbox{\hskip1pt$\sim$}}
     \raise1pt\hbox{$<$}}} 
\def\gsim{\mathrel{\rlap{\lower3pt\hbox{\hskip1pt$\sim$}}
     \raise1pt\hbox{$>$}}} 
\def\la{\langle}\def\ra{\rangle}

\def\cal{\it}

\begin{document}

\runauthor{Brown, Lee, Rho, \& Shuryak}

\begin{frontmatter}
\title{The $\bar q q$ Bound States and Instanton Molecules
 at $T \gsim T_C$}

\author[suny]{Gerald E. Brown,}
\author[pnu]{Chang-Hwan Lee,}
\author[saclay,kias,hanyang]{Mannque Rho,}
\author[suny]{Edward Shuryak}

\address[suny]{Department of Physics and Astronomy,\\
               State University of New York, Stony Brook, NY 11794, USA \\
               (\small E-mail: Ellen.Popenoe@sunysb.edu, shuryak@tonic.physics.sunysb.edu)}
\address[pnu]{Department of Physics and\\
Nuclear Physics \& Radiation Technology Institute (NuRI),\\
Pusan National University,
              Pusan 609-735, Korea\\ (E-mail: clee@pusan.ac.kr) }
\address[saclay]{Service de Physique Th\'eorique, CEA/DSM/SPhT. Unit\'e de
recherche associ\'ee au CNRS, CEA/Saclay, 91191 Gif-sur-Yvette
c\'edex, France\\ (E-mail: rho@spht.saclay.cea.fr)}
\address[kias]{School of Physics, Korea Institute for Advanced Study,
               Seoul 130-722, Korea}
\address[hanyang]{Department of Physics, Hanyang University, Seoul 133-791, Korea}
\renewcommand{\thefootnote}{\fnsymbol{footnote}}
\setcounter{footnote}{0}

\begin{abstract}

The main objective of this work is to explore the evolution in the
structure of the quark-anti-quark bound states in going down in
the chirally restored phase from the so-called ``zero binding
points" $T_{zb}$ to the full (unquenched)
QCD critical temperature $T_c$ at which
the Nambu-Goldstone and Wigner-Weyl modes meet. In doing this, we
adopt the idea recently introduced by Shuryak and Zahed for
charmed $\bar c c$, light-quark $\bar q q$ mesons $\pi, \sigma,
\rho, A_1$ and gluons that at $T_{zb}$,  the quark-anti-quark
scattering length goes through $\infty$ at which conformal
invariance is restored, thereby transforming the matter into a
near perfect fluid behaving hydrodynamically, as found at RHIC. We
show that the binding of these states is accomplished by the
combination of (i) the color Coulomb interaction, (ii) the
relativistic effects, and (iii) the interaction induced by the
instanton-anti-instanton molecules. The spin-spin forces turned
out to be small. While near $T_{zb}$ all mesons are large-size
nonrelativistic objects bound by Coulomb attraction, near $T_c$
they get much more tightly bound, with many-body collective
interactions becoming important and making the $\sigma$ and $\pi$
masses approach zero (in the chiral limit). The wave function at
the origin grows strongly with binding, and the near-local
four-Fermi interactions induced by the instanton molecules play an
increasingly more important role as the temperature moves downward
toward $T_c$.
\end{abstract}

\end{frontmatter}

\renewcommand{\thefootnote}{\arabic{footnote}}
\setcounter{footnote}{0}
\section{Introduction\label{intro}}
\subsection{$\bar q q$ bound states above $T_c$}

The concept that hadronic states may survive in the high
temperature phase of QCD, the quark-gluon plasma,
has been known for some time. In particular, it
 was explored by Brown et
al.\cite{BBP91,BJBP93}. The properties of (degenerate) $\pi$ and $\sigma$
resonances above $T_c$  in the context of the
NJL model was discussed earlier by Hatsuda
and Kunihiro\cite{Hatsuda_sigma},
 and in the instanton liquid model by Sch\"afer and Shuryak \cite{SS_survive}.
 Recently,
lattice calculations \cite{datta02,hatsuda2003} have shown
that, contrary to the original suggestion by Matsui and Satz  \cite{MS},
the lowest charmonium states $J/\psi,\eta_c$ remain bound well above $T_c$.
The estimates of the zero binding temperature for
charmonium
 $T_{J/\psi}$ is now limited to the interval
 $2T_c > T_{J/\psi} > 1.6 T_c$, where $T_c\approx 270\, MeV$
 is that for quenched QCD.
 Similar results for
 light quark mesons exist but are less quantitative at the moment.
However since the ``quasiparticle" masses close to $T_c$ are
large, they must be similar to  those for charmonium states.

  In the chiral limit~\footnote{In most of what follows, we will be ignoring
the effect of light-quark masses, unless mentioned otherwise. The
``quasiparticle mass" that we shall refer to in what follows is a
chirally invariant object called ``chiral mass."} all states above
 the chiral restoration go into
 chiral multiplets. For quark quasiparticles this is also true,
but although the chirality is conserved during their propagation,
they are not massless and move slowly near $T_c$ where their
``chiral mass'' $m=E(p\rightarrow 0)$ is large ($\sim 1$ GeV).

RHIC experiments have found that hot/dense matter at temperatures above the
critical value $T_c (unquenched)\approx 170 \, MeV$ is $not$ a weakly
interacting gas of quasiparticles, as was widely expected.
Instead, RHIC data have demonstrated the existence of  very robust
collective flow phenomena, well described by ideal hydrodynamics.
Most decisive in reaching this conclusion was the early
measurement of the elliptic flow which showed that equilibration
in the new state of matter above $T_c$ set in in a time $< 1$ fm/c
\cite{hydro2001}. Furthermore, the first viscosity estimates
\cite{Teaney2003} show  surprisingly low values, suggesting that
this matter is the most perfect liquid known. Indeed,  the ratio
of shear
  viscosity
coefficient to
  the entropy is only $\eta/s\sim 0.1$,  two orders of magnitude
  less than for water.
Furthermore, it is comparable to
predictions in the infinite coupling limit
(for $\cal N$=4 SUSY YM theory)  $\eta/s = 1/4\pi$ \cite{PSS}, perhaps the
lowest value possible.

Shuryak and Zahed\cite{shuryak2003} (hereafter referred to as SZ
whenever unambiguous) have recently connected these two issues
together. They have
 suggested that large rescattering cross sections apparently present
 in hot matter at RHIC
 are generated by resonances near the zero-binding lines.
 Indeed,  at the point of
zero binding  the scattering
length $a$ of the two constituents goes to $\infty$ and this
provides
low viscosity. This phenomenon is analogous to the  elliptic flow
observed in the expansion of trapped $^6$Li atoms rendered
possible by tuning the scattering length to very large values via
a Feshbach resonance~\cite{Li6}.

Near the zero-binding points, to be denoted by $T_{zb}$, introduced by SZ
the binding is small and thus
the description of the system can be simple and nonrelativistic.
The binding comes about chiefly
from the attractive Coulomb color electric field, as evidenced in
lattice gauge calculation of Karsch and
collaborators\cite{datta02,petreczky02}, and Asakawa and
Hatsuda\cite{hatsuda2003}, as we shall detail. The instanton
molecule interactions, which we describe below, are less important
 at these high temperatures ($T \sim 400$ MeV).
All this changes as one attempts (as we show below) to discuss
the more deeply bound states just above $T_c$ (unquenched).

In another  work \cite{SZ_CFT}, Shuryak and Zahed have also found
sets of highly relativistic bound light states  in the strongly coupled
 $\cal N$=4 supersymmetric Yang-Mills theory at finite
temperature (already mentioned above in respect to viscosity).
They suggested that the very strong Coulomb attraction can be
balanced by high angular momentum, producing light states with
masses $m\sim T$. Furthermore, the density of such states remains
constant  at arbitrarily large  coupling. They argued that in this
theory a transition from weak to strong coupling basically implies
a smooth transition from a gas of quasiparticles to a gas of
``dimers''\footnote{In the large number of
  color limit considered, those are dominated by bound
colored states, not colorless mesons that are important in the
regime we are considering.}, without a phase transition. This is
an important part of the overall emerging
 picture, relating strong coupling, viscosity and light bound states.

In this work we wish to construct the link between the chirally
broken state of hadronic matter below $T_c$ (unquenched) and the chirally
restored mesonic, glueball state above $T_c$.
Our objective  is to understand and to
work out in detail what exactly happens with hadronic states
at temperatures between
$T_c$ and $T_{zb}$.
 One important new point
 is that these chirally restored
hadrons are so small that the color charges are
locked into the hadrons at such short distances ($< 0.5$ fm)
that the Debye screening
is unimportant. This is strictly true at $T \gsim T_c$,
 where there is very little free charge.
In this temperature range the nonrelativistic treatment of SZ should
be changed to a relativistic one.

The relativistic current-current interaction, ultimately related with
the classical Ampere law,
 is about as important as the Coulomb one,
effectively doubling the attraction (see section
\ref{sec_coul_rel}).  We also found that the spin-spin forces
discussed in \ref{sec_spinspin} are truly negligible. In effect,
with the help of the instanton molecule interaction, one can get
the bound quark-anti-quark states down in energy, reaching the
massless $\sigma$ and $\pi$ at $T_c$,
 so that a smooth transition can be made with the chiral
breaking at $T<T_c$.

The nonperturbative interaction from the instanton molecules
becomes very important\footnote{Although we do not discuss $T<T_c$
in this work, we still mention that in this region the
instanton-induced effects seem to become dominant, see
\cite{SS_98} for review.}.  Let us remind the reader of the
history of the issue.
  The nonperturbative gluon condensate, contributing to
the dilatational charge or trace of the stress tensor
$T_{\mu\mu}=\epsilon-3p$,
 is not melted at $T_c$ (unquenched). 
 In fact more than half of the vacuum gluon condensate
 value remains at $T$ right above $T_c$.
 the hard glue or epoxy which
explicitly breaks scale invariance but is
 unconnected with hadronic masses. The rate at which the epoxy
is melted can be measured by lattice gauge simulations, and this
tells us the rate at which the instanton molecules are broken
up with increasing temperature. We will discuss this further in
 section \ref{sec_hard_glue}.

As argued by Ilgenfritz and Shuryak \cite{IS} (for further
references  see the review \cite{SS_98}), this phenomenon can be
explained by breaking of the instanton ensemble into instanton
molecules with zero topological charge. Such molecules generate a
new form of effective multi-fermion effective interactions similar
to the orignal NJL model, see details in section \ref{sec_mol}.
 Brown et al.\cite{bglr2003} (denoted as BGLR below) obtained the
interaction induced by the instanton molecules above $T_c$ by
continuing upwards the Nambu-Jona-Lasinio description from below
$T_c$.

  Our present discussion
of mesonic bound states  should not be confused with
quasi-hadronic states  found in early lattice
calculations\cite{detar85} for quarks and anti-quarks propagating
in the space-like direction. Their spectrum, known as ``screening
masses'' is generated mostly by ``dynamical confinement" of the
spatial Wilson loop which is a nonperturbative phenomenon seen via
the lattice calculations. Similar effects will be given here by
the instanton molecule interaction.  We will briefly discuss it in
section \ref{sec_screeningmasses}.


\subsection{Quasiparticles and their masses at $T>T_c$}
  \label{quasiparticles}

  It is well known that the expectation value of the
 Polyakov line  $|\langle L\rangle|$ goes to zero at $T_c$, indicating
an infinite quark mass below $T_c$; i.e., confinement. In the
deconfined phase we discuss this mass is finite, and as $T$ grows
it is expected to decrease to some minimal value, before growing
perturbatively as $M\sim gT$ at large $T$.

Chirality of the quarks is a good quantum number above $T_c$.
Now the chirally restored wave functions have good helicity, $+$
or $-$. Also it can be seen from eq.~(\ref{eq9}) that each wave
function has good chirality, $\vec\sigma\cdot\vec p\ \psi =\pm
p\ \psi$. The helicities and chiralities use up the 4 components in
the Dirac wave function. The fermion modes with equal (opposite)
 helicities and chiralities are called ``quark quasiparticles''
(``plasmino'' modes).

There are detailed results on their dispersion relations from pQCD.
 Specifically for quark quasiparticles,
 Weldon\cite{Weldon82} obtained the dispersion
relation
\be
p_0-|\vec p|=\frac{M_{th}^2}{|\vec p|}
\left(1+\left(1-\frac{p_0}{|\vec p|}\right)
\frac 12 \ln\left(\frac{p_0+|\vec p|}{p_0-|\vec p|}\right)\right)
\ee
which can be reasonably well approximated (to within 10\%) by $p_0^2\approx
M_{th}^2+{\vec p}^2$ where $M_{q}^2\equiv g^2 T^2/6$. The perturbative
gluon mass is $M_{g}^2\equiv g^2 T^2 
(N_c/2+N_f/6)/2$ \cite{Shu_JETP}.

If the exact dispersion relations for quasiparticles are known,
the 2-body interaction can be introduced by standard substitution
of frequency and momenta by the covariant derivatives with the
potentials.

  Unfortunately, the actual lattice data about the masses are very
fragmented, and are (to our knowledge)
only available from quenched calculations.
 The Polyakov loop
expectation value $|\langle L\rangle|(T)$ is plotted in Asakawa et
al.: at each $N_\tau$ it starts to deviate from zero at $T_c\sim
270$ MeV, the critical temperature for quenched calculations.
However, one has to renormalize the mass, removing the linear
divergent part of the point charge mass $\sim 1/a$ before
extracting any numbers, which needs more data.

 There are some results on quasiparticle dispersion relations from
 Petreczky et
al. \cite{petreczky02}, who find in (quenched) calculation
that they are all consistent with the usual dispersion relation
$\omega^2=k^2+m^2$ with the masses
$m_q/T=3.9\pm 0.2$, $m_g/T=3.4\pm 0.3$
or $m_q\sim 1.6$ GeV ,
$m_g\sim 1.4$ GeV for $T=1.5 T_c= 405$ MeV
where we take $T_c=270$ MeV. These quarks (and gluons) are
about as massive as charmed quarks.
Although obviously these masses will be reduced
in unquenched calculations, probably roughly by the ratio
of critical temperatures or so, it seems inevitable
that at $T\gsim T_c$ the quarks are so
massive~\footnote{We will argue for a somewhat lower values below,
which would perhaps arise from the unquenched calculations. Let
us  also remember that those are the ``chiral masses'' which do
not break chiral symmetry.} that their Boltzmann factors are
small,
inadequate  to furnish substantial presence of free quarks and gluons.
The light bound state would then be primary sources of the
pressure, as at low $T$.

Petreczky et al.\cite{petreczky02} omitted the first three values
(at lowest temperatures) from their fit of $\omega_q(p)$. Their
curve of $\omega_q^2 (p)/T^2$ intercepts the ordinate
at a value of $\sim 9$, which would give $\omega_q(p)\sim 3 T$,
or $\sim 1.2$ GeV. Furthermore, their $m_q$ are calculated with
bare quark masses which increases $m_q$, whereas our calculations
are in the chiral limit. At least for the moment, our use of
$m_q=1$ GeV seems not unreasonable.

In spite of such large uncertainties in the mass, we will be able
to carry out the  calculations of meson binding, expressing it in
units of $m_q(T)$, so that its precise value  won't matter. In
particularly, we will show that at $T\rightarrow T_c (unquenched)$ the 
binding of the pion and sigma mesons will go to $-2 m_q(T)$, making them
massless. The spin-spin and instanton molecule interaction will at
this point be less attractive for vector and axial mesons, which
will thus have a finite limit\footnote{Note however that the
Harada-Yamawaki vector manifestation (VM)~\cite{HY:PR} of chiral
symmetry predicts that the vector mesons also reach zero mass at
$T_c$ from below in accordance with the BR scaling. In a
forthcoming publication~\cite{BLR04}, an argument will be
developed to the effect that the vector mesons also go massless as
$T\rightarrow T_c$ from above.} of $M_{V,A}/m_q$ at $T\rightarrow
T_c$.


\subsection{The ``hard glue'' and instanton molecules at $T>T_c$}
  \label{sec_hard_glue}

 As we  already mentioned in the introduction,
the trace of the stress
tensor $T_{\mu\mu}=\epsilon-3p$,
 is not ``melted away'' at $T>T_c$, but is instead only a factor 2
smaller than at $T=0$.
Nice details on  this observable have been recently provided
 by David Miller\cite{miller03}, who used a set of the Bielefeld group
 lattice data and showed
 that  the hard glue is only melted by
$T\lsim 400$ MeV.

  The natural explanation of the existence of such ``hard glue'' or
``epoxy'' component, which survives chiral restoration and deconfinement,
 was given by the theory of instantons. It has been argued~\cite{IS} that chiral restoration
is not related to the instanton suppression at $T>T_c$,
 as was previously thought, but to their rearrangement, from
the quasi-random ensemble to that made of correlated
 instanton-anti-instanton pairs (or other clusters with net zero
 topological charge). Some
 details of the quantum/statistical mechanics of a $\bar I I$
molecules can be found in \cite{Shu_Vel}, the finite $T$
simulations of instanton ensemble has been shown
in \cite{SSV_95}, for other  references
see \cite{SS_98}.

 For a single molecule the contribution to the partition function
can be written as
\be Z=\int d\Omega d{\bar \Omega} |T_{\bar I I}|^{2N_f} exp(-S_g)\ee
where $\Omega,{\bar \Omega} $ are 12-dimensional collective variables
for $I,\bar I$, namely size, 4 positions and 7 $SU(3)_c$
angles\footnote{One of the angles conjugated to the Gell-Mann matrix
  $\lambda_8$,
 does not rotate the instanton solution since it has only 2 colors.}.
Here $T_{\bar I I}$ is the matrix element of the Dirac operator
between 2 zero modes, that of the instanton and anti-instanton.


  Instead of evaluating the coupling constant from first principles,
  one
can do in phenomenologically, in a NJL-like framework. Brown et al
\cite{bglr2003} (BGLR) showed that the scaled attraction which
breaks chiral symmetry below $T_c$ gives way to one only slightly
($\sim 6\%$) lower in which chiral symmetry is restored at $T_c$ (unquenched).
Whereas the Nambu Jona-Lasinio (NJL) below $T_c$ was connected to
the ``soft glue" part of (which may be associated with the
spontaneously generated part of~\footnote{An elaboration is
perhaps in order on the nature of symmetry breaking involved here
with the conformal invariance. As mentioned above and
elsewhere~\cite{bglr2003,lprv}, the trace anomaly of QCD
associated with the scale symmetry breaking has, roughly speaking,
two components: One ``soft" component which is locked to chiral
symmetry and the other ``hard" or ``epoxy" component which is not
directly tied to chiral symmetry. In fact in \cite{lprv}, it was
explicitly shown how in dense medium the soft component ``locks
onto" the property of chiral symmetry, with the melting of the
soft component corresponding to the melting of the quark
condensate. This is the notion used in the early discussion of BR
scaling~\cite{br91}. The simplest way to understand this
phenomenon in the case at hand is to consider the soft component
as resulting from an ``induced (or spontaneous) symmetry breaking"
and the hard component as an explicit symmetry breaking of
conformal symmetry. As is known since a long time~\cite{zumino},
conformal symmetry can be spontaneously broken {\it only if} it is
also explicitly broken. Thus we can identify the soft component of
the glue that melts across the phase transition as arising from
spontaneous breaking {\it in the presence of} of an explicit
breaking which remains intact across the phase boundary. This is
analogous to ``induced symmetry breaking (ISB)" of Lorentz
symmetry in the presence of chemical potential discussed in
\cite{langfeld}.}) the scale anomaly which gets restored with
chiral symmetry, at least half, if not more, of the glue -- which
we called ``epoxy"-- remained at $T_c$ and for some distance
above, being melted only gradually with increasing $T$.
The ``hardness" of this glue explains why $T_c (quenched)$ is
so much higher (by nearly 50 \%) than $T_c (unquenched)$.

\subsection{The ``screening'' masses and states }
\label{sec_screeningmasses}

  The issue was raised first by DeTar et al \cite{detar85} and
 explained  by Koch et al.\cite{KSBJ}. The  ``screening''
 states are formed due
to a specific nonperturbative phenomena in the magnetic sector of
high-$T$ QCD related with  ``dynamical confinement of spatial
Wilson loops". These phenomena exist at all $T>T_c$ and are best
explained \cite{KSBJ} in a ``funny space" in which the coordinates
$t$ and $z$ had been interchanged. In this way the new
``temperature" in the old $z$ direction was zero, whereas the new
$z$ is compactified  to $(\pi T)^{-1}$, essentially a dimensional
reduction.

In the Koch et al. work\cite{KSBJ},
in the old time (new $z$ direction), the Coulomb potential came
from a periodic array of charges, because of the periodicity in
this direction, which in the large-$T$ limit became that from a
wire
 \be
\Phi (r_{\perp}) = - \frac{g}{2\pi \beta} \; \ln(r_\perp/2 \beta)
 \ee
where $\beta=T^{-1}$. Such a potential had earlier been obtained in
QCD by Hansson \& Zahed \cite{Hasson92}.

 A linearly rising
potential from the space-like string tension was added, because it
was felt that confinement on top of the logarithmic attraction was
needed to hold the quark and anti-quark together. The
$-g^2(\vec\alpha_1\cdot\vec\alpha_2/r)$ interactions were,
however, in the two-dimensional $x, y$ directions in the disc
through current loops, and give about one eighth of the attraction
we shall find. Another eighth, as we outline below, comes from the
usual Coulomb attraction which goes as $-g^2/r$ at larger
distances, and which is sufficient to bind the quark and
anti-quark. The largest part of the binding around $T_c$ comes
from the instanton molecule interaction.

Koch et al.\cite{KSBJ} ignored the effective Coulomb interaction
(in the 2d form of that from a charged
wire) because of the dominance of the space-like
string tension.
Putting together
the above forces produced the two-dimensional $\pi$ and $\rho$ wave
functions in the $x,y$-plane. Since the $x$ and $y$ directions were
unchanged, on this plane we should recover these same wave functions
in a projected four-dimensional QCD calculation, but with increased
binding because of the additional interactions included.

The important feature of DeTar's spatially propagated states was that
quarks were still confined in colorless states above
$T_c$ (unquenched) \cite{detar85}.
However  in our work we discuss real propagating states, and
so  in the deconfined phase, $T>T_c$, those
can be  colored. The non-singlet $\bar q q$ states are only the
color octet ones, which is a channel with color repulsive force
and is obviously unbound: but
 states made of $qg$ and $gg$ type have colored channels with
an attraction and should exist. One more famous example
of colored bound states is the $qq$ Cooper pairs, leading to color
superconductivity phases \cite{super}. Although we could have discussed
all of them in the same way as we did the $\bar q q$ ones, we defer
this discussion to future works.
The 32 lowest $\bar q q$ states we consider here are,
however, colorless.\footnote{
In total, there are 32 degrees of freedom
of the $\bar q q$ bound states
which lie in mass well below the masses of quarks and
gluons, and are therefore the relevant variables for the
thermodynamics. This is somewhat fewer than the 37 equivalent
boson degrees of freedom
arising from perturbative quarks and gluons. This may be a partial
explanation of the $\sim 20\%$ lower number of degrees of freedom
than would result from quarks and gluons, found in lattice
calculations.
}
Thus in the temperature region up to $\sim 2 T_c$ (unquenched),
essentially up to the temperatures reached by RHIC, we have a dynamical
color confinement.

\section{Binding of the $\bar q q$ states}
\label{binding}
\subsection{The Coulomb interaction and the relativistic effects}
\label{sec_coul_rel}
At $T>T_c$ the charge is screened rather than confined \cite{Shu_JETP},
and so the potential has a general Debye form
\be V= {\alpha_s(r,T) \over r} exp\left(-{r\over R_D(T)}\right)\ee
(Note  that we use a (somewhat nonstandard)
 definition in which $\alpha_s$ absorbs the
  4/3 color factor.)
The general tenet of QCD tells us that the strength of the color
Coulomb should run. We know that perturbatively it should run as
\be
\alpha_s\sim\frac{1}{\log (Q/\Lambda_{\rm QCD})}
 \ee
with $\Lambda_{\rm QCD}\sim 0.25$ GeV. The issue is what
happens when the coupling is no longer small. In vacuum
we know that the electric field is ultimately confined to
a string, producing a linear potential.

In the plasma phase
this does not happen, and SZ assumed that the charge runs to
larger values, which may explain the weak binding at rather high
T we discussed in the
introduction.
Lattice results produce potentials which, when fitted in the form
$V(r)=-A \exp(-mr)+B$ with constant $A,B$ indeed
indicate\footnote{We thank Stefano Fortunato for useful discussion
of the issue and for the fits he provided to us. } that $A(T)$
grows above $T_c$ untill its maximum at $T=1.4\ T_c$, 
before starting to decrease
logarithmically at high $T$. The maximal value of the
average\footnote{The maximal value of the running coupling in SZ
was taken to be 1, but its value at the most relevant distances is
about 1/2 also.} coupling $max(A)\approx 1/2$. This is the value
which will keep charmonium bound, as found by Asakawa and Hatsuda,
up to $1.6 T_c$\cite{hatsuda2003}.\footnote{ We thank L\"oic
Grandchamps for this calculation.}

Running of the coupling is
not very important for this work in which
 we are mostly interested in deeply bound states related with
 short enough distances.
Therefore we will simply keep it as a non-running constant, selecting an
appropriate average value.

It is well known in the point charge Coulomb problem (QED) that
when $Z\alpha$ is increased and the total energy reaches zero there is a
singularity,
preventing
solutions  for larger $Z\alpha$.
 In the problem of the ``sparking of the vacuum" in
relativistic heavy ion collisions, the
 solution of the problem
was found by approximating the nuclei by
 a uniformly charged sphere; for a review of the history
see Rafelski et al.~\cite{rafelski78}. As a result of such
 regularization,
the bound electron level continues  past zero to $-m$,
at which point $e^+e^-$ production becomes possible
around the critical value of $Z_{cr}=169$.
 In short, the problem of the
point Coulomb charge could be taken care of by choosing a
distributed electric field which began from zero at the origin.

In QCD the charge at the origin is switched off
by asymptotic freedom, the coupling which runs to zero value at
the origin. A cloud of virtual fields making the charge
is thus ``empty inside''.
 We will  model a resulting potential
 for the color Coulomb interaction by simply
 setting the electric field equal to zero at $r=0$, letting it
decrease (increase in attraction) going outward~\footnote{Just at
$r=0$ the quark and anti-quark are on top of each other, so the
electric field is clearly zero. (Although the quark and anti-quark
are point particles, their wave functions will be distributed.)
For two rigid spheres, $V$ would take up the $1/r$ behavior only
at $2R$, but there will be some flattening. It will become clear
that our main conclusions are independent of these details.}. We
can most simply do this by choosing a charge distribution which is
constant out to $R$, the radius of the meson. If the original
$2m_q$ mass were to be lowered to zero by the color Coulomb
interaction and instanton molecule interaction, then the radius of
the final molecule will be \be R\simeq\frac{\hbar}{2 m_q}, \ee
although the rms radius will be substantially greater with the
instanton molecule interactions playing the main role around
$T_c$. \be
V &=& 
      - \alpha_s\frac{1}{2R}\left(3-\frac{r^2}{R^2}\right),
      \;\;\;\;\; r<R \nonumber\\
   &=&  - \alpha_s\frac{1}{r}, \;\;\;\;\; r>R.\label{9}
\label{potential} \ee This $V$ has the correct general
characteristics. As noted above, the electric field $\vec E$ must
be zero at $r=0$. It is also easy to see that $V$ must drop off as
$r^2/R^2$ as the two spheres corresponding to the quark and
anti-quark wave functions are pulled apart. Precisely where the
potential begins the $1/r$ behavior may well depend upon
polarization effects of the charge, the $+$ and $-$ charges
attracting each other, but it will be somewhere between $R$ and
$2R$, since the undisturbed wave functions of quark and anti-quark
cease to overlap here.

The $q\bar q$ system is similar to positronium in the equality of
masses of the two constituents. Since the main term value
is\footnote{To make comparison with QCD we should use $Z^2\alpha$
rather than $\alpha$ and remember the Casimir factor $4/3$.}
$m\alpha^2/4$, the $4$, rather than 2 in hydrogen, coming from the
reduced mass, one might think that the Coulomb, velocity-velocity
and other interactions would have to be attractive and 8 times
greater than this term value in order to bring the $2 m_q$ in
thermal masses to zero. However, this does not take into account
the increase in reduced mass with $\alpha$. Breit and Brown
\cite{Breit48} found an $\alpha^2/4$ increase in the reduced mass
with $\alpha$, or $25\%$ for $\alpha=1$, to that order. It should
be noted that in the Hund and Pilkuhn \cite{pilkuhn00}
prescription the reduced mass becomes $\mu=m_q^2/E$, which
increases as $E$ drops.

We first proceed to solve the Coulomb problem, noting that this gives us
the solution to compare with the quenched lattice gauge simulations,
which do not include quark loops.

Having laid out our procedure, we shall proceed with
approximations. First of all, we ignore spin effects in getting a
Klein-Gordon equation. The chirally restored one-body equation
which has now left-right mixing is
given by
 \be (p_0 +
\vec\alpha\cdot\vec p)\psi =0.\label{chidirac}
 \ee
Expressing $\psi$ in two-component wave functions $\Phi$ and
$\Psi$, one has \be
p_0\Phi &=& -\vec\sigma\cdot\vec p\Psi \nonumber\\
p_0\Psi &=& -\vec\sigma\cdot\vec p\Phi, \label{eq9} \ee giving the
chirally restored wave function on $\Psi$ \be \left(p_0
-\vec\sigma\cdot\vec p \frac{1}{p_0} \vec\sigma\cdot \vec p
\right)\Psi =0. \label{eq11}
 \ee
Here
 \be p_0=E_V=E+\alpha_s/r.
 \ee
Neglecting spin effects, $\vec\sigma\cdot\vec p$ commutes with
$p_0$, giving the Klein-Gordon equation $p_0^2 - \vec{p}^2=0$. We
now introduce the effective (thermal) mass, so that the equations
for quark and hole can be solved simultaneously following
\cite{pilkuhn00};
 \be
\left[ (\epsilon -V(r))^2-\mu^2 -\hat p^2 \right] \psi (r) =0
\label{klein} \ee where $\hat p$ is momentum operator, and the
reduced energy and mass are $ \epsilon = (E^2-m_1^2-m_2^2)/2E, \mu
= m_1 m_2 /E $ with $m_1=m_2=m_q$.

 Furthermore from eq.(\ref{eq9}),
 \be
\la \vec{\alpha}\ra=(\Psi^\dagger, \vec{\sigma}\Phi) +
(\Phi^\dagger,
\vec{\sigma}\Psi)=\frac{\vec{p}}{p_0}-\frac{i}{p_0}\la
[\vec{\sigma}\times\vec{p}]\ra.
 \ee
If $\vec{\sigma}$ is parallel to $\vec{p}$, as in states of good
helicity, the second term does
not contribute. From the chirally restored Dirac equation
(\ref{chidirac}), ignoring spin effects such as the spin-orbit
interaction which is zero in S-states we are considering, we find
$p_0^2=\vec{p}^2$.

Brown \cite{brown52} showed that in a stationary state the EM
interaction Hamiltonian between fermions is
 \be
H_{\rm int} =\frac{e^2}{r}\left(
1-\vec\alpha_1\cdot\vec\alpha_2\right),\label{hint}
 \ee
where the $\vec\alpha_{1,2}$ are the velocities. Applying
(\ref{hint}) to the chirally restored domain of QCD, we expect
 \be
H_{\rm int} &=& \frac{2\alpha_s}{r} \;\;\;\;\;\;\;\;{\rm for}\;\;
                            \vec\alpha_1\cdot\vec\alpha_2=-1\nonumber\\
            &=& 0  \;\;\;\;\;\;\;\;{\rm for}\;\;
  \vec\alpha_1\cdot\vec\alpha_2=+1
\label{eq14}
 \ee

\subsection{The spin-spin interaction}
\label{sec_spinspin}
  The nonrelativistic form of the spin-spin interaction,
in the delta-function form, may give an impression that it is
maximal at the smallest distances. However this is not true, as
becomes clear if the relativistic motion is included in full, and
in fact at $r\rightarrow 0$ it is suppressed. At large r,
when particle motion is slow, it is of course again suppressed,
thus contributing mostly at some intermediate distances.

This fact is clear already from the derivation of  the $1s$-state
hyperfine splitting Fermi-Breit due to hyperfine interaction in
hydrogen from 1930  \cite{fermi30} given by \be \delta H=\frac 23
(\vec\sigma\cdot\vec\mu) \int d^3r\frac{\psi^\dagger\psi}{r^2}
\frac{d}{dr}\frac{e}{E+e^2/r+m} \label{eq2.2} \ee Note the
complete denominator, which non-relativistically is just
substituted by m alone, but in fact contains the potential and is
singular at $r\rightarrow 0$. The derivative of the $e^2/r$ in the
denominator insured that the electric field was zero at $r=0$.
Here $\vec\sigma$ is the electron spin, $\vec\mu$ the proton
magnetic moment. In eq~(\ref{eq2.2}) the derivative can then be
turned around to act on $\psi^\dagger\psi$, and to order
$\alpha=1$ and with the $e^2/r$ neglected in the denominator, one
has
 \be \delta H\simeq
-\frac{8\pi}{3} (\vec\sigma\cdot\vec\mu)\frac{e}{2m} \psi^2(0),
\label{eq6}
 \ee
with $\psi$ taken to be the nonrelativistic $1s$ wave function to
lowest order in $\alpha$.

The hyperfine structure is obtained by letting the first $\vec p$
in eq.~(\ref{eq11}) act on the $p_0^{-1}$ and the second $\vec p$
go $\vec p +\sqrt{\alpha_s}\;\vec A$ with \be \vec A=\frac{\vec
\mu\times \vec r}{r^3} \ee with $\vec \mu$ the magnetic moment of
the anti-quark. One finds that the hyperfine structure is
\cite{fermi30} \be H_{\rm hfs} =\frac{1}{p_0^2} \sqrt{\alpha_s}\;
\vec\sigma \cdot [\vec E\times\vec A] \ee where $\vec E$ is the
color electric field. Thus, \be H_{\rm hfs}
=\frac{\sqrt{\alpha_s}\;|\vec E|}{p_0^2}
\left(\frac{\vec\sigma\cdot\vec\mu}{r^2}
-\frac{\vec\sigma\cdot\vec r\; \vec\mu\cdot\vec r}{r^4}\right)
=\frac 23 \frac{\sqrt{\alpha_s}\;|\vec E|}{p_0^2}
\frac{\vec\sigma\cdot\vec\mu}{r^2}. \ee where $|\vec
E|=2\alpha_s/r^2$. As in the hydrogen atom, the magnetic moments
of quarks and anti-quarks are
 \be
\mu_{q,\bar q} = \mp\frac{\sqrt{\alpha_s}}{p_0+m_{q,\bar q}}
 \ee
except that the Dirac mass $m_{q,\bar q}=0$ and $p_0$, in which
the potential is increased by a factor of 2 to take into account
the velocity-velocity interaction, is now \be p_0 =E+2
(\alpha_s/r) \ee for QCD so that in terms of the quark and
anti-quark magnetic moment operators~\footnote{The electric field
is denoted as $\vec{E}$ which should be distinguished from the
energy $E$.},
 \be
H_{\rm hfs}=-\frac 23 \frac{|\vec E|}{p_0 r^2}
(\vec\mu_q\cdot\vec\mu_{\bar q}).
 \ee
Of course, our $p_0$ for the chirally restored regime has
substantial $r$ dependence (whereas the $e/r$ in the hydrogen atom
is generally neglected, and $E+m$ is taken to be $2m$, so that
$\mu_e=-e/2m_e$). From Fig.~\ref{pot} it will be seen that (square
of) the wave function is large just where $\alpha_s/r$ is large.

For rough estimates we use averages. We see that, as in Table~\ref{tab1},
if $E$ is to be
brought down by $\sim 0.5 m_q$
for the $\sigma$ and $\pi$  by the
Coulomb interaction, then \be
2\langle \alpha_s/r\rangle \simeq  \frac 12  m_q\simeq \frac 14 p_0 \ee
so that with
$\alpha_s\sim 0.5 $, \be \langle r^{-1}\rangle \simeq \frac 12 m_q. \ee
We next see that this is consistent with the spin splitting forming a
fine structure of the two groups, the lower lying $\sigma$ and
$\pi$, and the slightly higher lying vectors and axial vectors.
Using our above estimates, we obtain
 \be \langle H_{\rm
hfs}\rangle &\simeq& \frac{1}{24}\;\frac{1}{16}
\vec\sigma_q\cdot\vec\sigma_{\bar q} m_q, \label{hhfs}
 \ee
so that for the $\sigma$ and $\pi$ where
$\vec\sigma_1\cdot\vec\sigma_2=-3$ we have \be \langle
H_{hfs}\rangle \sim -\frac{m_q}{128}, \ee the approximate equality
holding when $\alpha_s= 0.5$. Note that the hyperfine effect is
negligible for the $\alpha_s \sim 0.5$.
Although formally eq.~(\ref{hhfs}) looks like the hyperfine
structure in the chirally broken sector, it is really completely
different in makeup.

In our expression for $\langle H_{hfs}\rangle$ we have the $r$
dependence as $(p_0 r)^{-4} r^{-1}$ and $p_0 r=4$, basically
because the Coulomb interaction lowers the $\pi$ and $\sigma$ only
$1/4$ of the way to zero mass. This explains most of the smallness
of the spin-dependent interaction.

A recently renewed discussion of spin-spin
and spin-orbit interactions in a
 relativistic bound states has\ been made
Shuryak and Zahed \cite{SZ_spinorbit}, who derived their form for
both weak and  strong coupling
limits. Curiously enough, the spin-spin term changes sign between
these two limits: perhaps this is another reason why at intermediate
coupling considered in this work it happens to be so small.

\subsection{The effect of the instanton molecules}
\label{sec_mol}

The effective interaction is calculated as follows. The propagator
is written as \be S(x,y)=\sum_\lambda {\phi_\lambda^\dagger (x)
\phi_\lambda(y)
  \over \lambda+ im}\ee
where $\lambda, \phi_\lambda$ are eigenvalues and eigenfunctions
of the Dirac operator. If the
gauge field configuration is the anti-instanton-instantopn ($\bar
I I$) molecule, the 2 lowest eigenvalues are $\lambda=\pm |T_{\bar
I I}|$ and their eigenfunctions are simple combinations of zero
modes for instanton and anti-instanton $\phi_I\pm \phi_{\bar I}$.
The sum of those leads to \be S(x,y)={2\over |T_{\bar I I}|}
(\phi^\dagger_I(x)\phi_{\bar I}(y)+ \phi^\dagger_{\bar
  I}(x)\phi_{I}(y))\ee
This can be interpreted as follows: $\phi_I(x)$ is the amplitude
to go from point x to the instanton $I$, the other phi is the
amplitude to appear from the anti-instanton, and the factor in
front is the propagator between $I$ and $\bar I$.

\begin{figure}[h]
\begin{center}
\includegraphics[width=9cm]{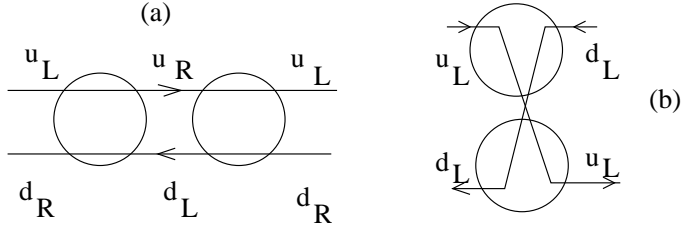}
\end{center}
\caption{
\label{4_ferm_diag}  The diagrams generating the effective Lagranfian
}
\end{figure}

 For propagation of 2 quarks, say of opposite flavors,
one can then draw the two diagrams shown in Fig.\ref{4_ferm_diag}.
The diagram (a) read from left to right has quarks of opposite
chirality, so it contributes to scalar and pseudoscalar mesons;
the diagram (b) has the same chirality of quark and anti-quark, so
it contributes to vector and axial vector channels. One can see
from the figure that in the former case both $\bar q q$ go into
the same instanton, while in the latter they have to go to the
opposite ones. As a result, the locality of the former vertex is
given by the instanton size $\rho$ and of the latter by
 the size of the molecule.

If all 4 points are far from the instanton and anti-instanton,
$\phi(x)\sim 1/x^3$ modulo a constant spinor. This power of the
distance corresponds to free propagator for
 a massless quark, $S_0=1/(2\pi^2 x^3)$, which should be
 ``amputated'',
leading to a constant vertex function, or effective Lagrangian
\cite{SSV_95}. The effective interaction between quarks
 is conveniently calculated by rearranging the exchange terms
 into a direct interaction. The resulting Fierz
symmetric Lagrangian reads \cite{SSV_95}
\be
\label{lmol}
 {\cal L}_{mol\,sym} &=&  G
     \left\{ \left[
     (\bar\psi\tau^a\psi)^2-(\bar\psi\tau^a\gamma_5\psi)^2
      \right]\right. \nonumber \\
     &&  - \;\,\frac{1}{4}\left. \left[
     (\bar\psi\tau^a\gamma_\mu\psi)^2+(\bar\psi\tau^a\gamma_\mu\gamma_5
     \psi)^2 \right] + 
     (\bar\psi\gamma_\mu\gamma_5\psi)^2 \right\}
      + {\cal L}_8,
\ee
with the last complicated term containing color-octet $\bar q q$ pairs only.
The coupling constant is now proportional to the density of molecules
\be
 G &=& \frac{2}{N_c^2} \int n(\rho_1,\rho_2)\,d\rho_1 d\rho_2\,
        \frac{1}{8T_{IA}^2}(2\pi\rho_1)^2(2\pi\rho_2)^2\, .
\label{gmol} \ee If the molecules are incompletely polarized in
the (Euclidean) time direction then all vector terms are modified
accordingly, because $ (\bar\psi\gamma_\mu\Gamma\psi)\sim \bar
z_\mu$, the only vector available.

We start by explaining the issue of the ``polarization'' of the
$\bar I I$ molecules. Let us
 put $I$ at the origin and in the standard SU(3) orientation
without rotation. Let us introduce the 4-d polar angle $\theta_4$:
the position vector of $\bar I$, called $\bar z_\mu$ is such that
  $\bar z_4=cos(\theta_4) \sqrt{\bar z_\mu^2}$. We will refer to
the $\theta_4=0$ case as molecules completely polarized in the time direction:
as discussed in \cite{SSV_95} this is where the maximum of the $Z$ is.
The very robust maximum corresponds to  ``half Matsubara box'',
$\bar z_4=1/(2T)$, and if $\theta_4$ is nonzero $Z$ is smaller,
basically
because of large distances in the $T_{\bar I I}$. Let us estimate the
effect of that, ignoring the gauge action,
\be Z\sim |T_{\bar I I}|^{2N_f} \sim \left({1 \over (2Tcos(\theta_4))^{-2}+\rho^2 }\right)^{2N_f}
\ee
which can be expanded into Gaussian form at small $\theta_4$. For 2
flavors and $T=T_c\approx 1/(4\rho)$, the root mean square polarization angle is
$\langle\theta_4\rangle\approx 0.55$. We will need below
\be \label{eqn_angles}
cos^2(\langle\theta_4\rangle)\approx .72 \hspace{2cm}
sin^2(\langle\theta_4\rangle)\approx .28
\label{eq33}
 \ee
 corresponding to $\langle\theta_4\rangle = 32^\circ$.

Let us remember that for vector particles propagating in matter
one can define in general 2 structures in the polarization operator,
the longitudinal and transverse ones $\Pi_{L,T}$, which are related to
1 longitudinal and 2 transverse modes. Those are related to
Cartesian components as follows
\be \Pi_{00}=\Pi_L, \hspace{1cm} \Pi_{0n}={\omega p_n \over \vec p^2}
\ \Pi_L \\
\Pi_{mn}= \left(\delta_{mn}-{p_m p_n \over  \vec p^2}\right)\ \Pi_T
+{\omega^2 p_m p_n\over  \vec p^2}\ \Pi_L
  \ee
where the Latin indices are 1-3 and Greek 0-3, $\omega=p_0$. This
polarization tensor satisfies the conservation law $p_\mu
\Pi_{\mu\nu}$, eliminating the 4-th component.

For complete polarization of molecules, $\theta_4=0$ and only
the zeroth component
(the longitudinal $\Pi_L$ component) is coupled,
while in general the coefficients include $sin^2$ and $cos^2$ of the angles
determined above (\ref{eqn_angles}) in  $\Pi_T$ and  $\Pi_L$, respectively.

This results in additional nonlocal correction factors, which  for
the scalar-pseudoscalar channels is \be \label{eq_overlap}
F_{nonlocal}^{S,PS}=\left|\int d^4x {\chi(x)\over
\chi(0)}|\phi_I(x)|^2 \right|^2 \ee and for the
vector-axial-vector ones \be
F_{nonlocal}^{V,A}=\left\langle\left|\int d^4x {\chi(x)\over
  \chi(0)}\phi_I(x)\phi_{\bar I}(x)\right|^2\right\rangle_{\theta_4}\ee
where the additional angular bracket in the latter case
comes about from averaging over the
molecular orientation relative to the time axes.
Note that in the former case, as well as in the latter for
$\theta_4=0$, the correction factors are 1,
due to the normalization condition $\int d^4x |\phi_I(x)|^2=1$, for weakly bound
(large size) states, for which the factor
$\chi(x)/,
  \chi(0)$ can be approximated by 1 and taken out of the integral.
We treat the ratios $(\chi (x)/\chi(0))^2=F$ as a vertex correction
in our calculations in the next section, and estimate their size
in the Appendix.


The effective interaction of light quarks,
 the Fierz symmetric instanton molecule Lagrangian of
Sch\"afer et al.\cite{SSV_95} in Minkowski space, is of the form
\be L_{IML} &=& G\left\{ (\bar\psi\tau^\alpha\psi)^2
     +(\bar\psi\tau^\alpha\gamma_5\psi)^2 \phantom{\frac 14}\right. \nonumber\\
    && \;\;\;\;\;
     +\left. \frac 14 \left[(\bar\psi\tau^\alpha\gamma_\mu\psi)^2
     -(\bar\psi\tau^\alpha\gamma_\mu\gamma_5\psi)^2
     \right]
     -(\bar\psi\gamma_\mu\gamma_5\psi)^2
     \right\}.
\label{eq1}
\ee
The value of $G$ at $T_c$ was obtained by BGLR~\cite{bglr2003} as
 \be G=3.83\; {\rm GeV}^{-2}. \label{eq2}
 \ee
Here $\tau^\alpha=(\tau,1)$ is a four-component vector. This
Lagrangian gives attraction in the $\sigma, \delta, \pi, \rho$ and
$A_1$  sectors. We will find that the mesons are bound by both the
color Coulomb and  the four-Fermi instanton molecule
interactions.~\footnote{As often the case in condensed matter and
particle physics, attractive four-Fermi interactions could -- via
anomalous dimensions -- figure crucially in phase transitions, so
may play a more important role near $T_c$. This may be
particularly relevant to other mesons than the $\pi$ and $\sigma$.
It is shown in \cite{BLR04} with a schematic model that the vector
mesons do undergo a phase change very similar to that of the $\pi$
and $\sigma$.}

We should explain how the chirally broken NJL is relayed into the
chirally restored NJL as $T$ goes upwards through $T_c$. The 't
Hooft instanton-driven interaction has been included in the
chirally broken NJL, and undoubtedly enters into the interactions
which bring the $m_\pi$ to zero and the $m_\sigma$ to $2 m_q$,
although their role relative to the other interactions is not
clear in this domain. As noted in BGLR\cite{bglr2003}, about half
of the total bag constant (conformal anomaly) coming from the
spontaneous breaking of chiral symmetry (the restoration of which
is associated with Brown/Rho scaling) is transferred from
that of the binding energy of the negative energy nucleons, which
are the relevant $T=0$, $\rho=0$ variables to the rearrangement of
the random instanton vacuum into the instanton molecules (which do
not break chiral symmetry) and the other half goes into melting
the soft glue as the nucleons loosen into constituent quarks.

\begin{figure}
\centerline{\epsfig{file=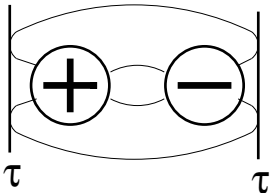,height=1in}} \caption{Looking
down on the instanton molecule close-packed around the
antiperiodic time direction at $T_c$. Here $\rho$ is the $\frac
13$ fm radius of the instanton or anti-instanton and
$\tau=1/4\rho\sim 150$ MeV. Just at $T_c$ the instanton molecule
is fully polarized in the $\tau$ direction. The lines going in and
out of the instanton and anti-instanton are the quark zero modes.
The outer quark zero modes are broken open to give the quark-quark
interaction which forms the instanton induced interaction. }
\label{figX}
\end{figure}

BGLR\cite{bglr2003} showed that NJL could be continued from the chirally
broken region up to the chirally restored one with the $\sim 6\%$ decrease
in the NJL coupling
\be
G &=& 4.08\; {\rm GeV}^{-2}\;\;\;\; {\rm chirally\  broken\  phase} \nonumber\\
G &=& 3.83\; {\rm GeV}^{-2}\;\;\;\; {\rm chirally\  restored\
phase}. \ee

We suggest that the fragility of chiral symmetry breaking as $T$
moves upwards to $T_c$ signifies the importance of the instanton
molecules in this region of $T$. With only an $\sim 6\%$ decrease
in $G$, the chiral restoration transition is effectuated.

Whereas in the chirally broken NJL, the vector interactions are
about the same size as the scalar ones, one can see from the
instanton molecule Lagrangian, eq.~(\ref{eq1}), that the $\rho$
and $A_1$ interactions are a factor of 4 smaller than the $\pi$
and $\sigma$ ones. However, as we have outlined, taking the
instanton molecule to be polarized along the time axis, the time
components of $\rho$ and $A_1$ interactions are built up a factor
of 4, at the expense of the spatial ones. Thus, in the classical
approximation, neglecting fluctuations in $\theta_4$, the vector
and axial vector modes would be degenerate with $\sigma$ and
$\pi$.


For the local 4-fermion interaction with the coupling constant $G$
the energy shift is given simply by
\be \delta E= G|\psi(0)|^2\ee
More generally, for the non-local interaction induced by molecules,
one should project on the (2-body) wave function of the bound state
$\Psi(x,y))=exp[iP_\mu(x+y)_\mu/2]\chi(x-y) $ where, for the stationary
state,
there is no dependence on relative time $x_0-y_0$.

The interaction is attractive in all
channels, so that if the Coulomb interaction does not bring the
meson mass all the way to zero, the instanton molecule
interactions will, in particular near the phase transition
temperature $T_c$, where the  wave functions are very
compact.

In the NJL-like instanton molecule Lagrangian eq.~(\ref{eq1}) the
interactions are four-point in nature. We can convert these to
an instanton molecule interaction by constructing a pseudo-potential
$V=C\delta (x)$ for the $\bar\psi\psi \rightarrow \bar \psi\psi$
scattering amplitude. Here $\langle\bar\psi \delta (x) \psi\rangle
=\bar\psi \psi (0)$. Therefore $C=G \bar\psi\psi (0)$ with $G$
the interaction of eq.~(\ref{eq2}), gives the strength of the
pseudo-potential.


Let us mention some estimates for the mass of the vector/axial
mesons. Keeping as above the thermal quark mass at $m_q=1 GeV$
and the $\alpha_s=0.5$ results from Table~\ref{tab1},
we will see that the Coulomb binding of $\sim 0.5$ GeV
plus that due to instanton
molecules $\sim 1$ GeV is able to make $\pi,\sigma$ mesons massless,
once we sum loops.

In the Appendix we estimated that the corrections for nonlocality
were roughly equal for $\pi$, $\sigma$ and $\rho$, $A_1$, so that the
masses of the latter scale as
$\cos^2 (\langle\theta_4\rangle)$ and
$\sin^2 (\langle\theta_4\rangle)$
relative to the $2 m_q\simeq 2$ GeV binding energy of the
$\pi$ and $\sigma$. From our estimates Eq.~(\ref{eq33}) we
see that the (predominantly) longitudinal $\rho$, $A_1$ masses
will come at $\sim 560$ MeV and the transverse $\rho$, $A$
masses at 1440 MeV; i.e., at roughly the free $\rho$ and $A_1$
masses, respectively.
However, the quasiparticle composed of coupled $\rho$ and $2\pi$
components (``rhosobar") may lie lower in energy when the
interaction is diagonalized.

\subsection{The resulting $\bar q q$ binding}
\label{sec_qqbin}

We first construct the bound states for $T\gsim T_c (unquenched)$,
at temperature close enough to $T_c$ so that we can take the
running coupling constants at $T=T_c+\epsilon$. The fact that we
are above $T_c$ is important, because the $\Lambda_{\rm\chi SB}
\sim 4\pi f_\pi\sim 1$ GeV which characterizes the broken symmetry
state below $T_c$ no longer sets the scale. Until we discover the
relevant variables above $T_c$ we are unable to find the scale
that sets $\alpha_s=\frac 43 \frac{g^2}{\hbar c}$, the color
Coulomb coupling constant.

Following SZ~\cite{shuryak2003}, we adopt quark-anti-quark bound
states to give the relevant unperturbed representation and, the
instanton molecule gas~\cite{SSV_95} as a convenient framework. In
particular, Adami et al.\cite{adami1991}, Koch and
Brown\cite{koch1993}, and BGLR~\cite{bglr2003} have shown that
$\gsim 50\%$ of the gluon condensate is not melted at $T=T_c$. The
assumption motivated by Ilgenfritz \& Shuryak~\cite{IS} is then
that the glue that is left rearranges itself into gluon molecules
around $T=T_c$, i.e., what BGLR call ``epoxy". We have
quantitatively determined couplings for the mesons in the
instanton molecule gas by extending the lower energy NJL in the
chiral symmetry breaking region up through $T_c$~\cite{bglr2003}.
We set these couplings in order to fit Miller's \cite{miller00}
lattice gauge results for the melting of the soft glue.

\begin{figure}
\centerline{\epsfig{file=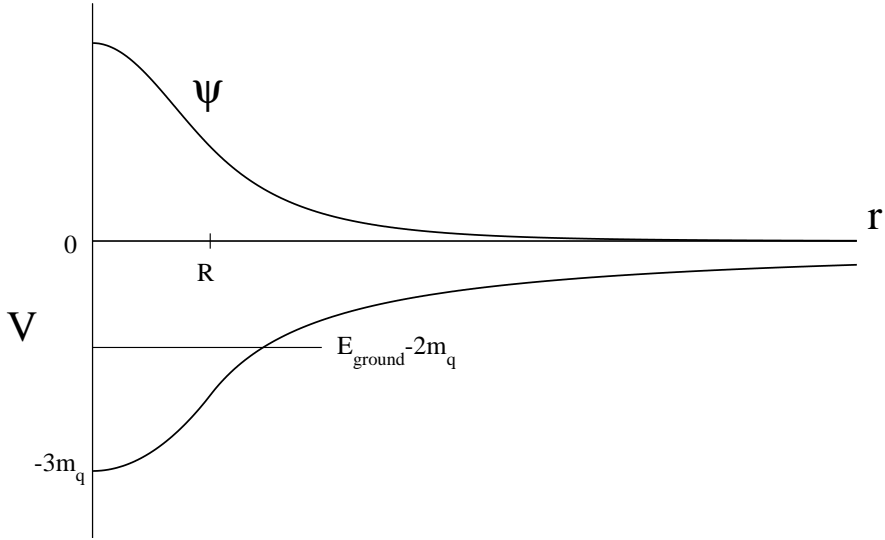,height=4.0in}}
\caption{The color Coulomb potential $V$ and
the corresponding wave function $\psi$ for
relativistic Klein-Gordon case. The interaction eq.~(\ref{potential})
with $R=\hbar/2 m_q$
was used.
The ground state energy with $\alpha_s=1$ corresponds to
$E_{\rm ground}=0.645\; m_q$. The minimum of the potential at the
origin is at $-3 m_q$ here.
}
\label{pot}
\end{figure}

In Fig.~\ref{pot} we show that if we choose $\alpha_s=1$
(effectively $\alpha_s=2$ by the doubling in Eq.~(\ref{eq14}))
as would be required to enter the strong coupling region
considered by Shuryak and Zahed\cite{SZ_CFT}
we bring the meson mass down by $-1.36\;
m_q$ from their unperturbed $2\; m_q$.
However, we switch to the region of $\alpha_s\sim 0.5$,
which is required by charmonium (intermediate coupling).
In Table~\ref{tab1} we summarize the Coulomb binding for a few choices
of $\alpha_s$.

 In the case of the instanton molecule interaction the
coupling constant $G=3.83$ GeV$^{-2}$ is dimensionful, so that
its contribution to the molecule energy scales as $G\; m_q^3$.
(Since we take $\alpha_s=0.5$ and will find that with inclusion
of the velocity-velocity interaction the effective $\alpha_s$ will
be 1, powers of $\alpha$ will not affect our answer. We will
use $m_q= 1$ GeV, essentially the lattice result for
$\frac 32 T_c$ and $3 T_c$\cite{petreczky02},
which works well in our schematic model.) Of course,
in QCD the Polyakov line goes to zero at $T_c$, indicating
an infinite quark mass below $T_c$; i.e., confinement.
Just at $T_c$ the logarithmically increasing confinement force
will not play much of a role because the dynamic confinement holds
the meson size to $\sim \hbar / m_q c$, or $\sim 0.2$ fm
with our assumption of $m_q =1$ GeV. (Later we shall see that
the rms radius is $\sim 0.3$ fm.) Since we normalize the instanton
molecule force, extrapolating it through $T_c (unquenched)$, and obtain the
color Coulomb force from charmonium, our $m_q$ is pretty well
determined. However, our $m_q=1$ GeV is for the unquenched system
and at a temperature where the instanton molecules play an important
role.

Given these caveats, we may still try to compare our Coulomb result
with the lowest peak of Asakawa et al.\cite{asakawa03} which is at
$\sim 2$ GeV for $T= 1.4 T_c \sim 0.38$ GeV and for Petreczky
at $\lsim 5 T\sim 2.030$ GeV for $T=1.5 T_c \sim 0.406$ GeV
where we used the Asakawa et al. $T_c$ (quenched). We wish to note that:
{\it (i)} These temperatures are in the region of temperatures
estimated to be reached at RHIC, just following the color glass phase
(which is estimated to last $\sim 1/3$ fm/c). Indeed, Kolb et al.
begin hydrodynamics at $T=360$ MeV. {\it (ii)} These are in the region
of temperatures estimated by SZ\cite{shuryak2003} to be those for
which bound mesons form.

We are unable to extend our consideration to higher temperatures,
where the situation may move towards the perturbative one, but we
believe that lattice calculations do support our scenario that
the QGP contains large component of bound mesons  from
$T\sim 170$ MeV up to $T\sim 400$ MeV.

\begin{table}
$$
\begin{array}{cccc}
\hline
\phantom{xxx} \alpha_s \phantom{xxx} &  \Delta E_{\rm Coulomb}\; {\rm [GeV]} &
\phantom{xxx} \sqrt{\langle r^2\rangle} \; {\rm [fm]} \phantom{xxx} &
\Delta E_{\rm 4-point} {\rm [GeV]} \\
\hline
 0.50 &  - 0.48 &  0.36 &  - 0.99 \\
 0.55 &  - 0.60 &  0.31 &  - 1.39 \\
 0.60 &  -0.71  &  0.28 & - 1.83 \\
 1.00 & - 1.36  &  0.14 & - 7.57 \\
\hline
\end{array}
$$
\caption{ Binding energies from color Coulomb interaction
and the corresponding rms radii for various $\alpha_s$
(effectively, $2\alpha_s$ including velocity-velocity interaction).
4-point interactions are calculated using the parameters
obtained from color Coulomb interaction.
}
\label{tab1}
\end{table}

\begin{figure}
\centerline{\epsfig{file=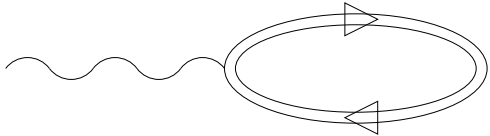,height=0.7in}}
\caption{Coulomb molecule. The wavy line on the left represents
the momentum transfer necessary to produce the molecule. The
double line denotes the Furry representation, i.e., 
the full propagator  in the Coulomb potential.} \label{fig3}
\end{figure}

For $\alpha_s=0.5$, which is the value required to bind charmonium
up through $T=1.6 T_c$, we find that the Coulomb interaction binds
the molecule by $\sim 0.5$ GeV, the instanton molecule interaction
by $\sim 1.5$ GeV. However, the finite size of the
$\psi^\dagger\psi$ of the instanton zero mode could cut the latter
down by an estimated $\sim 50\%$ (See Appendix). As in the usual
NJL, there will be higher order bubbles, which couple the Coulomb
and instanton molecule effects.\footnote{ Alternatively, the
$\delta$-function of the 4-point interaction should be included in
the Klein-Gordon equation, where it will change the energy.} We
draw the Coulomb molecule in Fig.~\ref{fig3}, where the double
lines denote the Furry representation (Coulomb eigenfunction for
quark and anti-quark in the molecule).

\begin{figure}
\centerline{\epsfig{file=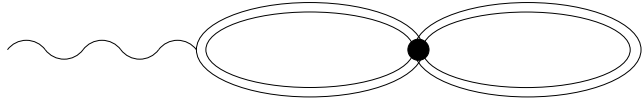,height=0.5in}}
\caption{The four-point instanton molecule interaction between Coulomb
eigenstates. The $(\bar\psi\psi)^2$ intersect at the thick point.
}
\label{fig4}
\end{figure}

\begin{figure}
\centerline{\epsfig{file=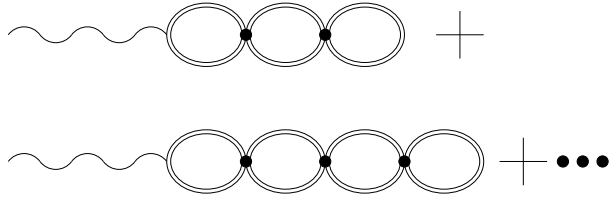,height=1.0in}}
\caption{Higher order effects of four-point interaction.}
\label{fig5}
\end{figure}

The four-point instanton molecule interaction is shown in Fig.~\ref{fig4}.
There will be higher-order effects as shown in Fig.~\ref{fig5}, of the
4-point interaction used in higher-order between Coulomb eigenstates
which always end in a 4-point interaction. The energy of the propagators
has been lowered from the 2 GeV of the two noninteracting quarks to
1.5 GeV by the Coulomb interaction. The series beginning with terms
in Figs.~\ref{fig3}$-$\ref{fig5} is
\be
\Delta E &=& -0.5 {\rm GeV} - 1 {\rm GeV}\; F
  -\frac{1 ({\rm GeV})^2 F^2}{1.5 {\rm GeV}}
  -\frac{1 ({\rm GeV})^3 F^3}{(1.5 {\rm GeV})^2} + \cdots
  \nonumber\\
&=& -0.5 {\rm GeV} -\frac{1 {\rm GeV} F}{1-\frac{1 {\rm GeV} F}{1.5 {\rm GeV}}}.
\label{eq43}
\ee
Now $\Delta E=-1.25$ GeV is accomplished for $F=0.5$,
a reasonable assumption from the
estimate in the Appendix.

Working in the Furry representation (including the Coulomb potential
exactly), we have a $-0.5$ GeV shift
already from the Coulomb wave functions.
This means that we must obtain $\Delta E=-1.5$ GeV to compensate
for the $2 m_q=2$ GeV, in order to bring the $\pi$ and $\sigma$
masses to zero. The four-point interaction is a constant, at
a given temperature, so this problem is just the extended
schematic model of nuclear vibrations (See Sec.~V of Brown
\cite{brown67}, where simple analytical solutions are given).

%

Our eq.~(\ref{eq43}) corresponds to the Tamm-Dancoff solution,
summing loops going only forward in time. If $\Delta E$ decreases
$-0.75$ GeV in this approximation, then when backward going
graphs 
are added, then $\Delta E$ will decrease by twice this amount \cite{brown67},
or the $-1.5$ GeV necessary to bring the $\pi$ and $\sigma$ energy
to zero. (There is an analogy between the treatment of the
spurion and the Goldstone mode in hadronic physics.)
Of course, forward and backward going loops are summed in the Bethe-Salpeter
equation to give the NJL in the broken symmetry sector, but
the actual summation is more complicated there, because the intermediate
state energies are not degenerate.

In detail, with our estimated $F= (0.75)^2$ from the Appendix and the
4-point energies from Table~\ref{tab1},
our $\pi$ and $\sigma$ excitations without inclusion of backwards going
graphs are brought down $58\%$ of the way from $-0.5$ GeV
to $-2$ GeV; i.e., slightly too far. We have not made the
adjustment down to $50\%$, because the uncertainties in our
estimate of $F$, etc., do not warrant greater accuracy.

\subsection{Comparison with Lattice Calculations}
\label{sec_lattice}

Lattice calculations propagating quarks and anti-quarks in the
spatial direction gave evidence of DeTar's dynamical confinement
\cite{Bernard92}. The dynamical confinement comes about from the
attractive $\alpha_s \vec\alpha_1\cdot\vec\alpha_2/r$, or
current-current interaction in our eq.~(\ref{hint}). This is the
same in the ``funny space", since the $x$ and $y$ directions, in
which the current loops lie, are as in the real space. Thus we
find from eq.~(\ref{hint}) that the introduction of the Coulomb
interaction doubles the effects from the current-current
interaction. The instanton molecule interaction gives a factor of
several times the current-current interaction. It should be
included in the unquenched lattice calculations.

It is instructive to examine the results of Bernard et
al.\cite{Bernard92} (reproduced in Fig.8.6 of Adami and Brown
\cite{adami93})
for two-dimensional wave functions
at $T=210$ and $T=350$ MeV of the chirally restored $\pi$ and
$\rho$. These wave functions are given in physical units. They
have been measured on the $(x,y)$-plane and we think of them as
projections onto this plane of the three-dimensional wave
function. The pion wave function at $T=210$ MeV is seen to drop
off exponentially, decreasing to 0.02 of its $r=0$ value by
$r=1.15$ fm. With a wave function $C\exp(-\kappa r)$, this
indicates a $\kappa$ of $3.4$ fm$^{-1}=0.68$ GeV. This gives an
rms radius for the two-dimensional wave function of $\sqrt{6}/2
\kappa$=1.8 GeV$^{-1} = 0.36$ fm. For the three-dimensional one we
would multiply by $\sqrt{3/2}\simeq 1.22$; thus one arrives at an
rms radius of $\sim 0.44$ fm. This is not much larger than the
0.36 fm for our chosen $\alpha =0.5$, and it should not be because
of the relatively small role  played
by the Coulomb term.

It can be seen that the $\rho$-meson wave function does not drop
exponentially until $2-3$ fm, showing it to be larger in extent
than the pion. It then appears to decrease somewhat more slowly
than the pion wave function, although the errors are such that
there is uncertainty in this.
The true quasiparticle above $T_c$ may, however, be a linear
combination of $\rho$ and 2 pion states.

The instanton molecule model, as the DeTar dynamical confinement,
involves an analytical continuation from imaginary to real time,
although it is quite simple in the former. The SZ~\cite{shuryak2003}
Coulomb-bound gas of mesons is formulated directly in real time. It
is simple and straightforward. We view it to be useful to give our
alternative instanton molecule formulation, because it enables us
to make contact with the lattice calculation, especially with
those of the gluon condensate.

Hadronic spectral functions above the QCD phase transition have been
calculated in quenched lattice gauge simulations by Asakawa, Hatsuda,
\& Nakahara \cite{asakawa03} and by Petreczky \cite{petreczky03}.

In the quenched approximation, the wavy line in Fig.~\ref{fig3}
can be interpreted as the source current which creates a valence
quark and anti-quark. These are allowed to exchange any number of
gluons, so it is appropriate to let them exchange any number of
Coulomb interactions, as well as develop thermal masses; i.e., the
Coulomb problem is that which we outlined.

We differ in detail from SZ\cite{shuryak2003}, in that we consider
the 32 normal modes of $\bar q q$ states, in which the quark and
anti-quark, or quark and hole, have opposite helicities so as to
benefit from the current-current interaction. SZ have the bound
states of quark and anti-quark, or quark and hole, and,
separately, those of gluons, the glueballs. Because of the larger
Casimir operator for the gluons, these bind at a higher
temperature than the quarks, so there would be two regions of
temperature in which the molecules break up. The number of degrees
of freedom, 40 quarks and 16 gluons, once they break up, would be
different. Presumably they would have the same $q\bar q$ bound
states for $T\sim T_c$ as we, but, in addition, would have the
glueballs.

As noted by SZ, in the region where the molecules break up, the quark
velocities are small, going to zero at zero binding energy. This means
that the velocity-velocity interaction will be unimportant and the
additional 32 degrees of freedom, disfavored at lower temperatures by
the velocity-velocity interaction, will become important, doubling
the degrees of freedom.

However, the quenched approximation would not include the quark
and anti-quark loops of the instanton molecule interaction. Thus,
at $T=T_c$ we would have molecules of energy $1.5$ GeV, since our
Coulomb interaction gives a 0.5 MeV binding on thermal quark and
anti-quark masses, each of 1 GeV. (In fact, these masses were
measured at $1.5 T_c (quenched)$, and need not be the same at $T_c$.
Furthermore, we need to mention that our $T_c$ is that for the
unquenched calculations, $T_c=170-175$ MeV, because we necessarily
have a situation with quarks and anti-quarks, especially instanton
zero modes.)

\begin{figure}
\centerline{\epsfig{file=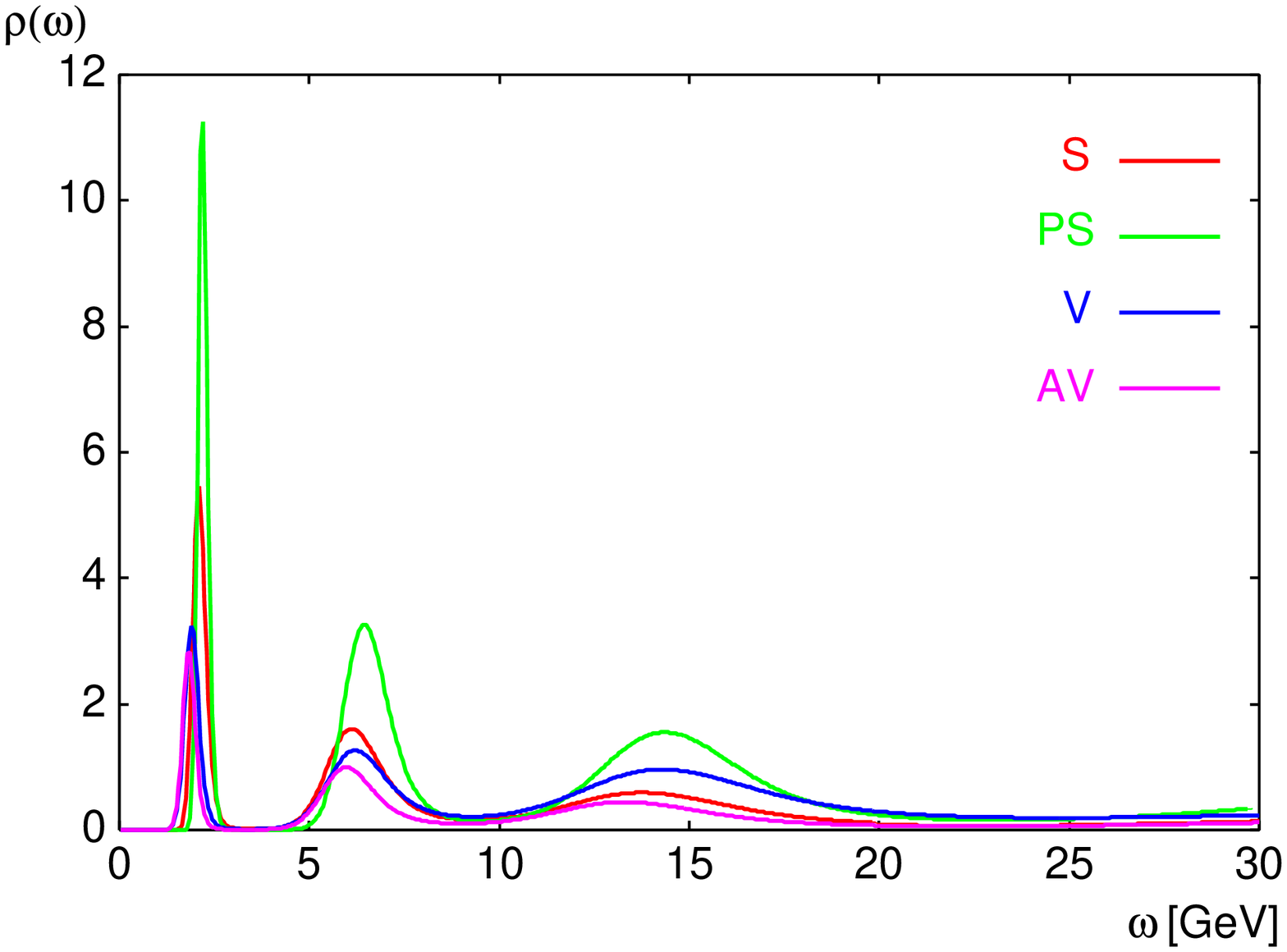,height=2in}
\epsfig{file=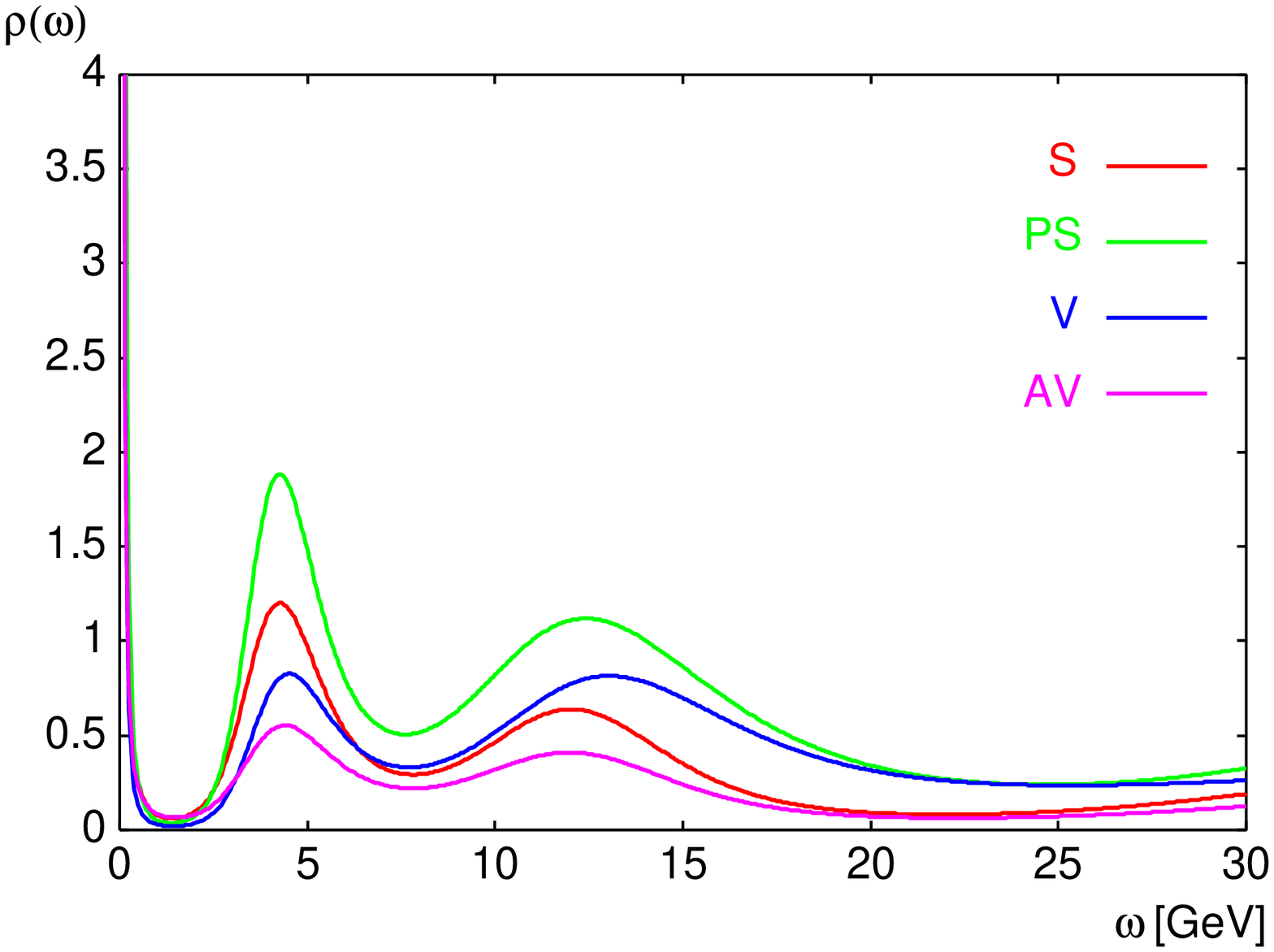,height=2in}}
\caption{Spectral functions of Asakawa et al.\cite{asakawa03}.
Left panel: for $N_{\tau}=54$ ($T\simeq 1.4 T_c$).
Right panel: for $N_{\tau}=40$ ($T\simeq 1.9 T_c$).}
\label{figZ}
\end{figure}

In Fig.~\ref{figZ} we show Asakawa \& Hatsuda\cite{hatsuda2003}
spectral functions
for (quenched) temperatures of $1.4 T_c\sim 380$ MeV and
$1.9 T_c\sim 515$ MeV. The lower temperature is essentially that
reached at RHIC following the color glass  phase
and the second temperature is higher, but probably
not in the perturbative regime because the lattice calculations by
Petreczky et al.\cite{petreczky02} give the quark mass at $3T_c$ as
roughly the same as at $\frac 32 T_c$, not increasing as $gT$.

The narrow 2 GeV peak supports our identification of this
temperature ($T\sim 380$ MeV) being where the molecules break up,
because the particle velocities will begin from zero at
breakup.\footnote{In the LGS the zero-momentum state of the
quark-anti-quark pair is projected out.} In this case, the 2 GeV
reproduces the sum of quark and anti-quark masses. We note the
meson degeneracy is consistent with our negligible effects from
spin and chiral restoration of the $\pi$ and $\sigma$ and vector
and axial vectors.

%
%

The $1.9 T_c\sim 515$ MeV data which should be well above the
temperature at which the molecules break up does show definite
thermal effects. As noted earlier, the lattice calculations give
approximately the same $m_q$ at $3 T_c$ as at $(3/2) T_c$, so we
can take $m_q$ to be constant in this region of energies. Each of
the $\sim 1$ GeV quark and hole in quark and anti-quark will have
a thermal energy of $\sim 1.1$ GeV, midway between the $\frac 32
T$ nonrelativistic thermal energy and the $3T$ relativistic one.
Therefore, it costs $\sim 4.2$ GeV to produce a quark particle and
hole or quark and anti-quark at this temperature. The half width
of the peak is roughly consistent with the size of the proposed
thermal energy. 

At low $\omega$ we probably see Landau damping in the $1.9 T_c$
results, although at $1.4 T_c$ there is no such sign. This further
supports our belief that the quark and anti-quarks are essentially
at rest there.

By rescaling the lattice calculation Peter Petreczky (private
communication) has shown that the peaks higher than the
first one are lattice artifact.

We see that in the SZ\cite{shuryak2003} model, one can understand
not only in general terms, but also many of the details of the
Asakawa and Hatsuda lattice calculations.

\section{Conclusions}

Shuryak and Zahed have discussed the formation of the mesonic
bound states at higher temperatures, well above $T_c$. They
pointed out that in the formation of the bound state, or any one
of the molecular excited states, the quark-quark scattering length
becomes infinite, similarly for the more strongly bound
gluon-gluon states. In this way the nearly instantaneous
equilibration found by RHIC can be explained. As we explained in
the last section, lattice calculations seem to support the
scenario of nearly bound scalar, pseudoscalar, vector and
axial-vector excitations at $\sim 2 T_c (unquenched)$  
($\sim 1.5$ times the quenched $T_c$).

In this work we are able to construct a smooth transition from the
chirally broken to the chirally restored sector in terms of
continuity in the masses of the $\sigma$ and $\pi$ mesons,
vanishing at $T\rightarrow T_c$. In doing so we had to include
relativistic effects. One of them -- the velocity-velocity term
related to Ampere law for the interacting currents -- nearly
doubles the effective coupling. The spin-spin term happens to be
very small. The crucial part of strong binding in our picture of
$\bar q q$ mesons (or molecules) is the quasi-local interaction
due to instanton molecules (the ``hard glue''). We found that the
tight binding of these mesons near $T_c$ enhances the wave
function at the origin, and gives us additional understanding of
the nonperturbative hard glue (epoxy) which is preserved at
$T>T_c$.

Thus, we believe that the material formed in RHIC was at a
temperature where most of it is made of
chirally restored mesons.  Certainly this is not
the weakly coupled quark-gluon plasma expected at high $T$.

Finally, in this paper we have focused on quantum mechanical
binding effects in the vicinity of the critical temperature $T_c$
coming down from above.  Nice continuity in
the spectra of the light-quark hadrons -- e.g., the pions and the
$\sigma$ -- across the phase boundary should also hold
for other excitations such
as the vector mesons $\rho,\omega, A_1$ which lie slightly above $\pi$
and $\sigma$ because of quantum corrections. Since going below
$T_c$ from above involves a symmetry change from Wigner-Weyl to
Nambu-Goldstone, there is a phase transition and to address this
issue, it would be necessary to treat the four-fermi interactions
more carefully than in the pseudo-potential approximation adopted
here. As briefly noted, the true quasiparticles in the $\rho$-channel in the
many-body medium may be linear combinations of the $\rho$ found here
and $2\pi$ states.

\section*{Acknowledgments}
We would like to thank Santo Fortunato, Peter Kolb, Peter Petreczky,
Madappa
Prakash and Ismail Zahed for many extremely useful discussions.
GEB and ES were partially supported by the US Department of Energy
under Grant No. DE-FG02-88ER40388. CHL is supported by Korea
Research Foundation Grant (KRF-2002-070-C00027).


\appendix
\section{Appendix: Finite Size Corrections}

The instanton is $\sim 1/3$ fm in radius, and the rms radius of
the molecule is about the same for $\alpha_s=0.5$, so there will
be a correction for the finite size of the instanton zero modes.
Their wave functions are \be
\psi_{ZM}=\frac{\rho/2\pi}{(x^2+\rho^2)^{3/2}} \ee with $\rho$ the
instanton radius. Now \be \bar\psi\psi\propto
\frac{1}{(x^2+\rho^2)^3}. \ee Integrating over time \be \int d\tau
\frac{1}{(x^2+\rho^2)^{3}}  &\sim & \frac{1}{(r^2+\rho^2)^{5/2}}
\propto \exp \left[-\frac 52 \ln
\left(\frac{r^2}{\rho^2}+1\right)\right]
\nonumber\\
 &\approx & \exp\left[-\frac 52 \frac{r^2}{\rho^2} \right].
\ee
Thus, we see that $\bar\psi\psi$ is sharply peaked, mostly lying within
a radius $r\sim\sqrt{2/5}\;\rho$
or a volume of $\left(\sqrt{2/5}\; \rho\right)^3$, or $\gsim 25\%$ of the
instanton molecule.
This nonlocality will spread the initially forward peaked
$\bar q q$ wave function over a volume of about $1/4$ of the
instanton, so we estimate $F=(0.75)^2$ in Sec.~\ref{sec_qqbin}.

Our above estimate holds for the effect of nonlocality in the
scalar and pseudoscalar mesons, where its ratio to the size of the
instanton is calculated. In the vector and axial-vector the
ratio should be to that of the entire molecule, but the nonlocality
is also over the whole molecule, and the wave function, due to
less binding, will not be so forward peaked as in the $\pi$ and
$\sigma$. Thus, we use the same $F=(0.75)^2$.

The possible considerable error in this estimate is important
in determining the role of the vectors and axial-vectors in the
thermodynamics of the system, but not in our main purpose of
constructing continuity in the $\pi$ and $\sigma$ masses
across $T_c$.

Ultimately the vector and axial-vector masses may be quantitatively
determined for $T\gsim T_c$ if the  lattice gauge simulation
of the Asakawa and Hatsuda type (see Sec.~\ref{sec_lattice})
are extended to unquenched ones.


\end{document}